\documentclass[doublecol]{epl2}

\title{Plane symmetric model with constant deceleration parameter}
\shorttitle{Plane symmetric model with constant DP} 

\author{Vijay Singh\inst{1}\footnote{gtrcosmo@gmail.com} \and Aroonkumar Beesham\inst{1,2}\footnote{abeesham@yahoo.com}}
\shortauthor{V. Singh and A. Beesham}

\institute{
  \inst{1} Department of Mathematical Sciences,\\
            University of Zululand,
              Private Bag X1001 \\
              Kwa-Dlangezwa, 3886,  \\
              South Africa \\
  \inst{2} Faculty of Natural Sciences, \\Mangosuthu University of Technology, P O Box 12363, \\Jacobs 4052, \\South Africa
}

\abstract{
A plane symmetric Bianchi I model is considered with constant deceleration parameter, $q=\alpha-1$, where $\alpha\geq0$. The model with $\alpha=0$ violates the NEC throughout the evolution, and hence provides a physically unrealistic scenario. The model with $\alpha\neq0$ obeys the NEC and WEC at late times, which shows that the models in this case can render a physical realistic cosmological scenario, though for a restricted period of time. It is also shown that the physical and kinematical behaviour of both models remain similar to an LRS Bianchi I model.}

\begin{document}

\maketitle

To find cosmological solutions in general relativity (GR), one normally assumes an equation of state for the matter, and then proceeds to find the scale factor.  A reverse approach is always possible, i.e., finding the matter required to give a desired geometrical  state. Following this reverse procedure, Berman \cite{BermanNC1983} proposed a special law of variation for the Hubble parameter
\begin{equation}
  H=\beta a^{-\alpha},
\end{equation}
where $\alpha\geq0$ and $\beta>0$ are constants. Here $a(t)$ is an average scale factor. The deceleration parameter, $q=-1-\dot H/H^2$ (a dot denotes the ordinary derivative with respect to cosmic time $t$), for the above law returns the constant value
\begin{equation}
  q=\alpha-1.
\end{equation}
Consequently, Eq. (1) can also be referred as the law of constant deceleration parameter. The models with $\alpha>1$ describe decelerating universes, whereas the models with $\alpha<1$ correspond to accelerating universes.

The Hubble parameter is defined as $H=\dot a/a$, therefore, Eq. (1) yields \cite{BermanGomideGRG1988}
\begin{equation}
a(t)=\left\{
  \begin{array}{ll}
     a_0e^{\beta t}, & \hbox{$\alpha=0$;} \\
    \left[\alpha\left(\beta t+t_0\right)\right]^\frac{1}{\alpha}, & \hbox{$\alpha\neq0$,}
  \end{array}
\right.
\end{equation}
where $a_0$ and $t_0$ are integration constants. Thus, Berman's law is a combination of  the de Sitter and power-law expansions. The models with $\alpha\neq0$ have singular origin while those with $\alpha=0$ have non-singular origin.

Substituting Eq. (3) into Eq. (1), one has
\begin{equation}
H=\left\{
  \begin{array}{ll}
    \beta, & \hbox{$\alpha=0$;} \\
    \beta\left[\alpha(\beta t+t_0)\right]^{-1}, & \hbox{$\alpha\neq0$.}
  \end{array}
\right.
\end{equation}
Many authors have solved Einstein's equations using (1)--(4) in GR as well as in some of the others theories of gravity (see Ref. \cite{SinghBeeshamGRG2019} and references therein for a complete list of earlier works). These works were carried out in homogenous and isotropic space-times. Maharaj and Naidoo \cite{MaharajNaidooASS1993} applied  Berman's law to anisotropic models also. They exemplified the case of the Bianchi I metric. Following them, many authors have  considered this law to obtain solutions in various anisotropic space-time models (see Ref. \cite{SinghBeeshamGRG2019} and references therein). Assuming the same law in Ref. \cite{SinghBeeshamGRG2019}, we have explored the physical and geometrical properties of an LRS Bianchi I space-time in the framework of GR. In a Cartesian coordinate system, the line-elements of the LRS Bianchi I space-time and plane symmetric space-time are indistinguishable. Consequently, the solutions in both models remain the same. Both models behave similarly both physically and geometrically. However, an interchange of directional expansion rate is obvious due to the formulation of the metric in both models. The purpose of this letter is to show this phenomenon explicitly.

\section{Model and the field equations}
The spatially homogenous and anisotropic plane symmetric Bianchi I space-time metric is given by
\begin{equation}
ds^{2} =dt^{2}-A^2(dx^2+B^2)dy^2+dz^2,
\end{equation}
where $A$ and $B$ are the scale factors, and are functions of the cosmic time $t$.
The average scale factor and average Hubble parameter, respectively, are defined as
\begin{eqnarray}
  a&=&(A^2B)^{\frac{1}{3}},\\
  H&=&\frac{1}{3}\left(2\frac{\dot A}{A}+\frac{\dot B}{B}\right).
\end{eqnarray}
The Einstein field equations read as
\begin{equation}
R_{\mu\nu}-\frac{1}{2}R g_{\mu\nu}=T_{\mu\nu},
\end{equation}
where the system of units $8\pi G=1=c$ is used.

The energy-momentum tensor of a perfect fluid is given as
\begin{equation}
  T_{\mu\nu}=(\rho+p)u_\mu u_\nu-pg_{\mu\nu},
\end{equation}
where $\rho$ is the energy density and $p$ is the thermodynamical pressure of the matter.

Equations (8), for the metric (5), with the consideration of the energy-momentum tensor (9), yield
\begin{eqnarray}
  \left(\frac{\dot A}{A}\right)^2+2\frac{\dot A\dot B}{A B}&=&\rho,\\
2\frac{\ddot A}{ A}+\left(\frac{\dot A}{A}\right)^2&=&-p,\\
\frac{\ddot A}{A}+\frac{\ddot B}{B}+\frac{\dot A \dot B}{AB}&=&-p.
\end{eqnarray}
From Eqns. (9) and (10), the condition of isotropy of pressure is
\begin{equation}
\frac{\dot A}{A} -\frac{\dot B}{B}=\frac{k}{A^2B},
\end{equation}
where $k$ is a constant of integration.

\section{Solutions with $\alpha=0$}
\label{sec:2}
Equations (7) and (13) by the use of Eq. (4), give
\begin{eqnarray}
  A&=&c_1e^{\beta  t+\frac{k }{9  \beta }e^{-3 \beta  t}}\\
B&=&c_1e^{\beta  t-\frac{2 k }{9  \beta }e^{-3 \beta  t}},
\end{eqnarray}
where $c_1$ is an integration constant and $a_0$ is taken to be unity without loss of generality.

From Eqns. (10)-(12), the energy density and pressure are
\begin{eqnarray}
  \rho&=&3 \beta ^2-\frac{1}{3} k^2 e^{-6 \beta  t},\\
  p&=&-3 \beta ^2-\frac{1}{3} k^2 e^{-6 \beta  t},
\end{eqnarray}
which are identical to the one in the LRS Bianchi I model \cite{SinghBeeshamGRG2019}.

 To study the NEC\footnote{$\rho+p\geq0$}, we require
\begin{equation}
  \rho+p=-\frac{2}{3}k^2 e^{-6 \beta  t},
\end{equation}
which remains negative always. Hence, a constant expansion rate ($H=\beta$) or a constant deceleration parameter ($q=-1$) leads to a violation of the NEC. The NEC is an essential requirement for any physical scenario to be realistic.A violation of this condition leads to the violation of other energy conditions as well. Hence, the model with $\alpha=0$ provides an unrealistic cosmological scenario. In other words, the assumption of a constant deceleration parameter or a constant expansion rate is not consistent with a plane symmetric space-time model. The same is also true for an LRS Bianchi I model. However, in Ref. \cite{SinghBeeshamGRG2019}, at a cost of a violation of the NEC and the WEC, we have advocated that the model can explain  late time cosmic acceleration or even eternal inflation. The motivation fo the arguments is the positivity of the energy density for a time
\begin{equation}
  t\geq\frac{1}{\beta}\ln\left(\frac{k^2}{9\beta^2}\right)^\frac{1}{6}.
\end{equation}
A hypothetical exotic matter which violates the NEC is well known as a phantom scalar field. Therefore, in Ref. \cite{SinghBeeshamGRG2019}, we have also considered a scalar field (quintessence or phantom) model, and showed that the solutions are compatible with a phantom scalar field. Thus, all the results and discussion related to the physical behavior presented in Ref. \cite{SinghBeeshamGRG2019} will be true in the present model also.

\subsection{Geometrical behavior ($\alpha=0$)}
\label{sec:4}
The expansion rates along the $x$, $y$, and $z -$axes are
\begin{eqnarray}
  H_x=\frac{\dot B}{B}&=& \beta-\frac{1}{3} k  e^{-3 \beta t},\\
  H_y=H_z=\frac{\dot A}{A}&=&\beta+\frac{2}{3} k  e^{-3 \beta t}.
\end{eqnarray}
The expansion scalar, $\theta$ and the shear scalar, $\sigma$ are
\begin{equation}
 \theta=\frac{\dot A}{A}+2\frac{\dot B}{B}=3\beta,
\end{equation}
\begin{equation}
\sigma^2=\frac{1}{3}\left(\frac{\dot A}{A}-\frac{\dot B}{B}\right)^2=\frac{1}{3} k^2 e^{-6 \beta  t}.
\end{equation}
The anisotropy parameter is
\begin{equation}
  \mathcal{A}=\frac{1}{3}\sum_{i=1}^3\left(\frac{H_i-H}{H}\right)^2=\frac{2 k^2 e^{-6 \beta  t}}{9 \beta ^2},
\end{equation}
where $H_i$ ($i=1,2,3$) represents the directional Hubble parameters in the directions $x$, $y$, $z$ respectively. On comparing these geometrical parameters with the one in an LRS Bianchi I model \cite{SinghBeeshamGRG2019}, we see that only the directional expansion rates are interchanged, but all other parameters remain identical in both models. Hence, both models evolve similarly kinematically as well.

\section{Solutions with $\alpha\neq0$}

Equations (7) and (13) by the use of Eq. (4), give
\begin{eqnarray}
  A&=&\left\{
  \begin{array}{ll}
     c_2t^{\frac{1}{3}-\frac{k }{9 \beta}}, & \hbox{$\alpha=3$;} \\
    c_2t^\frac{1}{\alpha} e^{\frac{k  \alpha t} {9-3 \alpha}(\beta \alpha t)^{-\frac{3}{\alpha}}}, & \hbox{$\alpha\neq3$,}
  \end{array}
\right.\\
B&=&\left\{
  \begin{array}{ll}
     c_2t^{\frac{1}{3}+\frac{2k }{9 \beta}}, & \hbox{$\alpha=3$;} \\
    c_2t^\frac{1}{\alpha} e^{\frac{2 k  \alpha t} {3 (\alpha-3)}(\beta \alpha t)^{-\frac{3}{\alpha}}}, & \hbox{$\alpha\neq3$,}
  \end{array}
\right.
\end{eqnarray}
where $c_2$ is an integration constant and we have taken $t_0=0$ to shift the origin to  $t=0$.

\subsection{Model with $\alpha\neq3$}

The energy density and pressure become
\begin{eqnarray}
 \rho&=& \frac{3}{\alpha ^2 t^2}-\frac{1}{3} k^2 (\alpha  \beta  t)^{-\frac{6}{\alpha }},\\
p&=&\frac{2 \alpha -3}{\alpha ^2 t^2}-\frac{1}{3} k^2 (\alpha  \beta  t)^{-\frac{6}{\alpha }},
\end{eqnarray}
which are also identical to those one in the LRS Bianchi I model. The NEC in this case is obeyed at late times. Therefore, this model represents a realistic cosmological scenario during late evolution. Hence, all the results of Ref. \cite{SinghBeeshamGRG2019} will also be valid for this model.

\subsection{Geometrical behavior ($\alpha\neq0$)}

The rates of the expansion along with the  $x$, $y$, and $z -$axes give
\begin{eqnarray}
  H_x=\frac{1}{\alpha  t}-\frac{1}{3} k (\alpha  \beta  t)^{-\frac{3}{\alpha }},\\
  H_y=H_z=\frac{1}{\alpha  t}+\frac{2}{3} k (\alpha  \beta  t)^{-\frac{3}{\alpha }}.
\end{eqnarray}
The expansion scalar, $\theta$ and the shear scalar, $\sigma$ become
\begin{eqnarray}
 \theta&=&\frac{3}{\alpha  t},\\
\sigma^2&=&\frac{1}{3} k^2 (\alpha  \beta  t)^{-\frac{6}{\alpha }}.
\end{eqnarray}
The anisotropic parameter is
\begin{equation}
  \mathcal{A}=\frac{2}{9} \alpha ^2 k^2 t^2 (\alpha  \beta  t)^{-\frac{6}{\alpha }}.
\end{equation}
Again, we see that only the directional expansion rates are interchanged, but the other geometrical parameters remain identical to the one in the LRS Bianchi I model. Hence, both models behave alike geometrically in this case also.

Similarly, one can see that the model for $\alpha=3$ also produces the results indistinguishable from an LRS Bianchi I model. The expansion rate will be interchanged, but the physical and geometrical behavior for both models remain the same.

 \section{Discussion and conclusion}
A plane symmetric Bianchi I model is considered with constant deceleration parameter, $q=\alpha-1$. It has been shown that the physical and geometrical behavior of the model are similar to an LRS Bianchi I model. The model with $\alpha=0$ violates the NEC, which does not offer a physically realistic scenario. In other words, the assumption $q=-1$ does not fit with the plane anisotropic or LRS Bianchi I model. However, the solutions can be defended by the use of  phantom matter. The model can explain a late time accelerating phase or an eternal inflating universe as we have studied recently \cite{SinghBeeshamGRG2019}. On the other hand, in the model with $\alpha\neq0$ both  the NEC and WEC hold for late times. Therefore, the models with different values of $\alpha$ can describe accelerating and decelerating phases of the cosmological evolution for a restricted period of time. The geometrical properties in both cases ($\alpha=0$ and $\alpha\neq0$) remain similar to an LRS Binachi I model \cite{SinghBeeshamGRG2019},  except an interchange in the directional expansion rate which is obvious due to the mathematical structure of the space-times.

\acknowledgments{This work is based on the research supported wholly/in part by the National Research Foundation of South Africa (Grant Numbers: 118511).} Dr V Singh is grateful to the University of Zululand for a postdoc fellowship and for facilities provided.

\end{document}